\documentclass[aps,prx,twocolumn,reprint,amssymb,floats,superscriptaddress]{revtex4-2}

\usepackage{graphicx}  
\usepackage{multirow}
\usepackage{svg}
\usepackage{physics}
\usepackage{tikz}

\usepackage[normalem]{ulem}
\usepackage{hyperref}

\usepackage{longtable}
\usepackage[T1]{fontenc}
\usepackage{dcolumn}   

\usepackage{bm}        
\usepackage{amsfonts}  
\usepackage{amsmath}   
\usepackage{amssymb}   

\newcommand{\re}[1]{\mathrm{Re}\left\{#1\right\}}

\newcommand{\pwisein}{\left\{ \begin{array}{ll}}
\newcommand{\pwiseout}{\end{array}\right.}

\usepackage{mathtools}

\begin{document}

\title{Approximate encoding of quantum states using shallow circuits}
\author{Matan Ben-Dov}
\affiliation{Department of Physics, Bar-Ilan University, 52900 Ramat Gan, Israel}
\affiliation{Center for Quantum Entanglement Science and Technology, Bar-Ilan University, 52900 Ramat Gan, Israel}
\author{David Shnaiderov}
\affiliation{Department of Physics, Bar-Ilan University, 52900 Ramat Gan, Israel}
\affiliation{Center for Quantum Entanglement Science and Technology, Bar-Ilan University, 52900 Ramat Gan, Israel}
\author{Adi Makmal}
\affiliation{Center for Quantum Entanglement Science and Technology, Bar-Ilan University, 52900 Ramat Gan, Israel}
\affiliation{Faculty of engineering, Bar-Ilan University, 52900 Ramat Gan, Israel}

\author{Emanuele G. Dalla Torre}
\affiliation{Department of Physics, Bar-Ilan University, 52900 Ramat Gan, Israel}
\affiliation{Center for Quantum Entanglement Science and Technology, Bar-Ilan University, 52900 Ramat Gan, Israel}

\date{\today}

\begin{abstract}  
A common requirement of quantum simulations and algorithms is the preparation of complex states through sequences of 2-qubit gates. For a generic quantum state, the number of gates grows exponentially with the  number of qubits,  becoming unfeasible on near-term quantum devices. Here, we aim at creating an approximate encoding of the target state using a limited number of gates. As a first step, we consider a quantum state that is efficiently represented classically, such as a one-dimensional matrix product state. Using tensor network techniques, we develop an optimization algorithm that approaches the optimal implementation for a fixed number of gates. Our algorithm runs efficiently on classical computers and requires a polynomial number of iterations only.  We demonstrate the feasibility of our approach by comparing optimal and suboptimal circuits on real devices. We, next, consider the implementation of the proposed optimization algorithm directly on a quantum computer and overcome inherent barren plateaus by employing a local cost function rather than a global one. By simulating realistic shot noise, we verify that the number of required measurements scales polynomially with the number of qubits. Our work offers a universal method to prepare target states using local gates and represents a significant improvement over known strategies. 
\end{abstract}

\pacs{47.15.-x (whats that?)}

\maketitle 


\section{Introduction}

Quantum state preparations is the task of generating a circuit that prepares a user-defined target state. This task is the first step of several quantum algorithms, such as linear solvers  \cite{duan2020survey}, quantum machine learning \cite{schuld2018information,leymann2020bitter}, and quantum recommendation systems \cite{kerenidis2016quantum}, where classical information is encoded in the initial state of the circuit. Similarly, in the quantum simulation of nonequilibrium systems, the initial state is often a complex state, for example, corresponding to the ground state of a many-body Hamiltonian \cite{calabrese2006time,polkovnikov2011colloquium,mitra2018quantum,paeckel2019time}. 
Because a generic state of $N$ qubits is defined uniquely by $2^N$ independent amplitudes, encoding such state requires an exponentially large number of gates \cite{barenco1995elementary,long2001efficient,grover2002creating,mottonen2004transformation,shende2004minimal,shende2004quantum,vartiainen2004efficient,shende2006synthesis,plesch2011quantum,aaronson2015read,sun2021asymptotically}. 

In real devices, the maximal number of gates per circuit is heavily limited by noise and decoherence. Quantum computing practitioners often assume that the maximal achievable circuit depth scales linearly with the number of qubits (see, e.g., the definition of quantum volume by IBM \cite{cross2019validating}). To match this requirement, it is necessary to heavily reduce the total number of gates used in state encoding. One possible strategy is to consider segments of the circuit, searching for a smaller circuit that performs the same operation, for example by scanning lists of known circuit identities \cite{pointing2021optimizing,Epping2022, xu2022quartz}. This method deterministically reduces the number of gates, but is limited to previously studied problems and does not apply to generic multi-qubit states. Another common method is to perform a blind search in a variational space of quantum circuits \cite{ Khatri_2019,sharma2020noise,Jones_2022,meister2022exploring,birtea2022constraint, wiersema2022optimizing,nakaji2022approximate, tepaske2022optimal}, trying to incrementally approach the target state. The problem with this approach is that due to the large number of variational parameters, the optimization algorithm is, again, limited to a small number of qubits only.


In this paper, we develop an efficient method to obtain an approximate encoder, given a fixed number of gates. We, first, consider quantum states that can be efficiently represented classically, such as matrix product states (MPS) \cite{MPS_perez2007,schollwock2011density,Or_s_2014,Bridgeman_2017,TNRan_2020}. In this case, the optimization algorithm can be realized efficiently on a classical computer, using resources that scale polynomially with the number of gates. We exemplify our protocol for both MPSs with low entanglement and random MPSs and demonstrate a significant improvement with respect to known techniques. Next, we consider the realization of our algorithm on a quantum computer, where the optimization protocol is affected by inherent shot noise. This error source leads to barren plateaus that limit the applicability of optimization algorithms. We overcome this limitation by introducing an improved version of the algorithm, based on local cost functions, whose resources requirements scale polynomially with the number of gates.

\section{background: layer-by-layer encoding of MPS states}
%
%
In the first part of our work we encode quantum states that can be efficiently expressed using classical means, and in particular, using open-boundary matrix-product states (MPS). In this representation, a one-dimensional state of $N$ qubits is expressed in terms of $2N$ matrices, $\lbrace A_i^{(0)}\rbrace_{i=1}^N$ and $\lbrace A_i^{(1)}\rbrace_{i=1}^N$ according to 
\begin{equation}
    \ket{\psi} =  \sum_{j_k \in \{0,1\}}A_{1}^{(j_1)} A_{2}^{(j_2)}\dots A_{N}^{(j_N)}\ket{j_1 ... j_N}.\label{eq:MPS}
\end{equation}
Here, $\ket{ j_1 ... j_N}$ is the computational basis and we assumed open boundary conditions, such that the first and last matrices are, respectively, row- and column- vectors. The maximal dimension of $A_i$ is denoted by $\chi$, and is known as the bond dimension. Here, we will consider realistic situations where the bond dimension is in the intermediate regime $1\ll\chi\ll 2^{N/2}$.  As we will see below, $\chi$ determines the resources required to realize the MPS on a quantum computer in an exact manner. By fixing $\chi$, one obtains a variational family of states with maximal entanglement entropy $S_{\rm max}=\log_2\chi$. For this reason, MPSs offer a faithful description of the ground state of one-dimensional gapped Hamiltonians, whose entanglement entropy follows an area law and does not grow with $N$ \cite{ hastings2007area,eisert2008area,eisert2010colloquium,eisert2013entanglement,arad2013area}. Similarly, MPS are useful in quantum machine learning tasks, such as ansatz structures \cite{Huggins_2019} and ansatz pertaining \cite{dborin2022matrix}.  

From Eq.~(\ref{eq:MPS}) it is possible to directly derive an (inefficient) quantum circuit that encodes the MPS \cite{Sch_n_2005, Ran_2020}. To achieve this goal, one needs to consider each $A_i$ as a rank-3 tensor, acting on vectors spanned by the left bond-index space, and transforming them into higher-dimension vectors spanned by the right bond-index and the physical index \(j_i\) \footnote{The $A_i$ matrices are defined up to a gauge transformation, which can be fixed by selecting a center of orthogonality and demanding that all tensors on its left and right sides, respectively satisfy $\protect\sum{A}_{j_i}^{b_{i-1} b_{i}} {{A}_{j_i}^{b_{i-1}' b_{i}}}^*= \delta^{b_{i-1}'}_{b_{i-1}}$ and $\protect\sum{{A}_{j_i}^{b_{i-1} b_{i}} {{A}_{j_i}^{b_{i-1} b_{i}'}}^*} = \delta^{b_{i}'}_{b_{i}}$. The gauge fixing procedure ensures that the tensors are injective and preserve orthogonality.}. This tensor can be transformed into a unitary gate by adding an extra input index, while preserving its isometric property \footnote{This is analogous to adding columns to an orthogonal rectangular matrix until it becomes a square unitary matrix.}. By combining the resulting $N$ gates in sequence, one obtains an exact circuit that creates the desired MPS. The problem is that each gate acts on one physical qubit and $\log_2(\chi)$ ancilla qubits associated with the bond index. Expressing this gate in terms of native 2-qubit gates requires a complex  circuit with depth exponential in the number of ancilla qubits \cite{barenco1995elementary,LI_2013, Plesch2011,iten2016, bergholm2005}, which would be too noisy for near-term devices (See section \ref{sec:classi_opt} for more details). 


\begin{figure}
	\centering
	\includegraphics[width=\columnwidth]{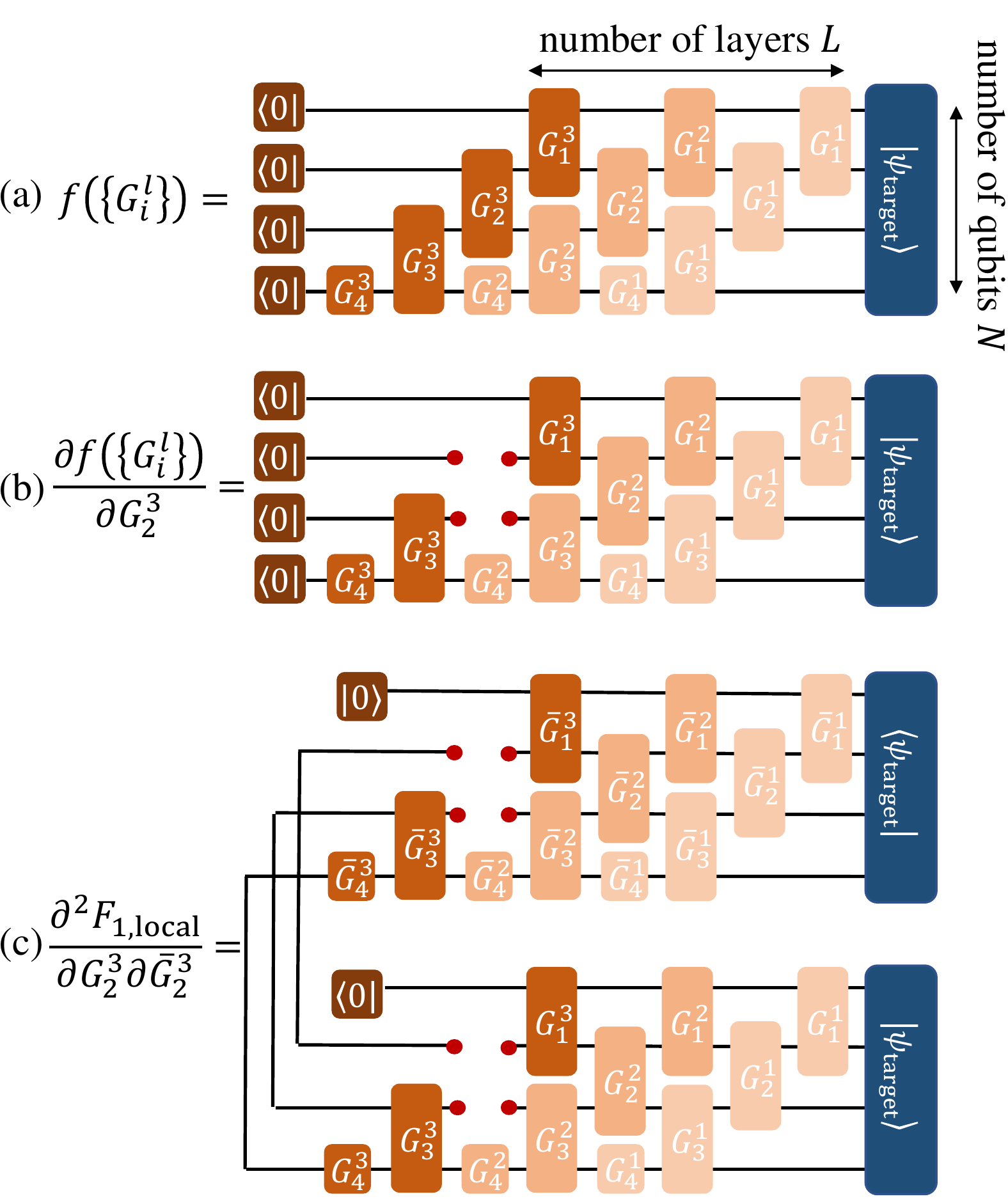}
	\caption{(a) Tensor diagram of the overlap between the target state $\ket{\psi_{\rm target}}$ and a staircase circuit with $N=4$ qubits and $L=3$ layers. (b) Gradient of the overlap used in the  global optimization protocol. (c) Gradient of the fidelity used to optimize the local cost function. The red dots denote the indices of the resulting tensor.}
	\label{fig:tens_cost_func}
\end{figure}

A natural question is whether it is possible to obtain an {\it approximate} description of the state using a smaller number of two-qubit gates. A solution to this problem was proposed by Ran \cite{Ran_2020}, following a layer-by-layer approach: (i) The first layer is obtained by simply truncating the bond dimension of the MPS to $\chi = 2$ \footnote{The truncation of the bond dimension is a common procedure, obtained by contracting and re-separating each neighbour pairs of tensors trough a singular-value decomposition (SVD) and keeping the two largest values.}. As explained above, the resulting state can be easily translated into a sequence of $N$ 2-qubit gates. (ii) The next layer is generated by applying the inverse of the first layer to the target state and truncating the resulting MPS to $\chi=2$, and so on and so forth. This protocol generates a staircase circuit of 2-qubit gates acting on neighboring pairs of qubits only, which we denote by $\{G_i^l\}_{i=1,\dots,N}^{l=1,\dots,L}$, where $L$  is the number of layers, see Fig.~\ref{fig:tens_cost_func}(a)  \footnote{From a mathematical perspective, the gates $\lbrace G_i^l\rbrace$ can be described either as unitary operators or as tensors, where the each approach has its own advantages and disadvantages. In particular, the tensor approach is convenient from a computational point of view, but misses the definitions of input and output channels and its associated unitarity.}. This circuit is suitable for all near-term quantum computers, including processors with limited connectivity such as superconducting circuits, and will be used throughout this work.


The key advantage of the layer-by-layer approach is that each step of the calculation can be performed while keeping the state in an MPS form and using tensor contractions. In this way, the complexity of computing each layer grows linearly with the number of qubits, in contrast to a naive calculation whose complexity grows exponentially with the number of qubits. Note that the classical computational complexity scales quadratically with the bond dimension, which grows by a factor of 2 for each additional layer, i.e.\ exponentially with the number of layers. Hence, the maximal number of layers is bounded by the classical resources available and is typically set to 20 or less. To achieve a larger number of layers, it is necessary to perform a truncation of the MPS representation of the resulting state to a fixed bond dimension $\chi_{\rm max}$, at the probable cost of deteriorating the overlap with the target state. At present, however, this is not an important limitation, because near-term devices are anyway limited to a small number of layers.



\begin{figure}
    \centering
    \includegraphics[width=\columnwidth]{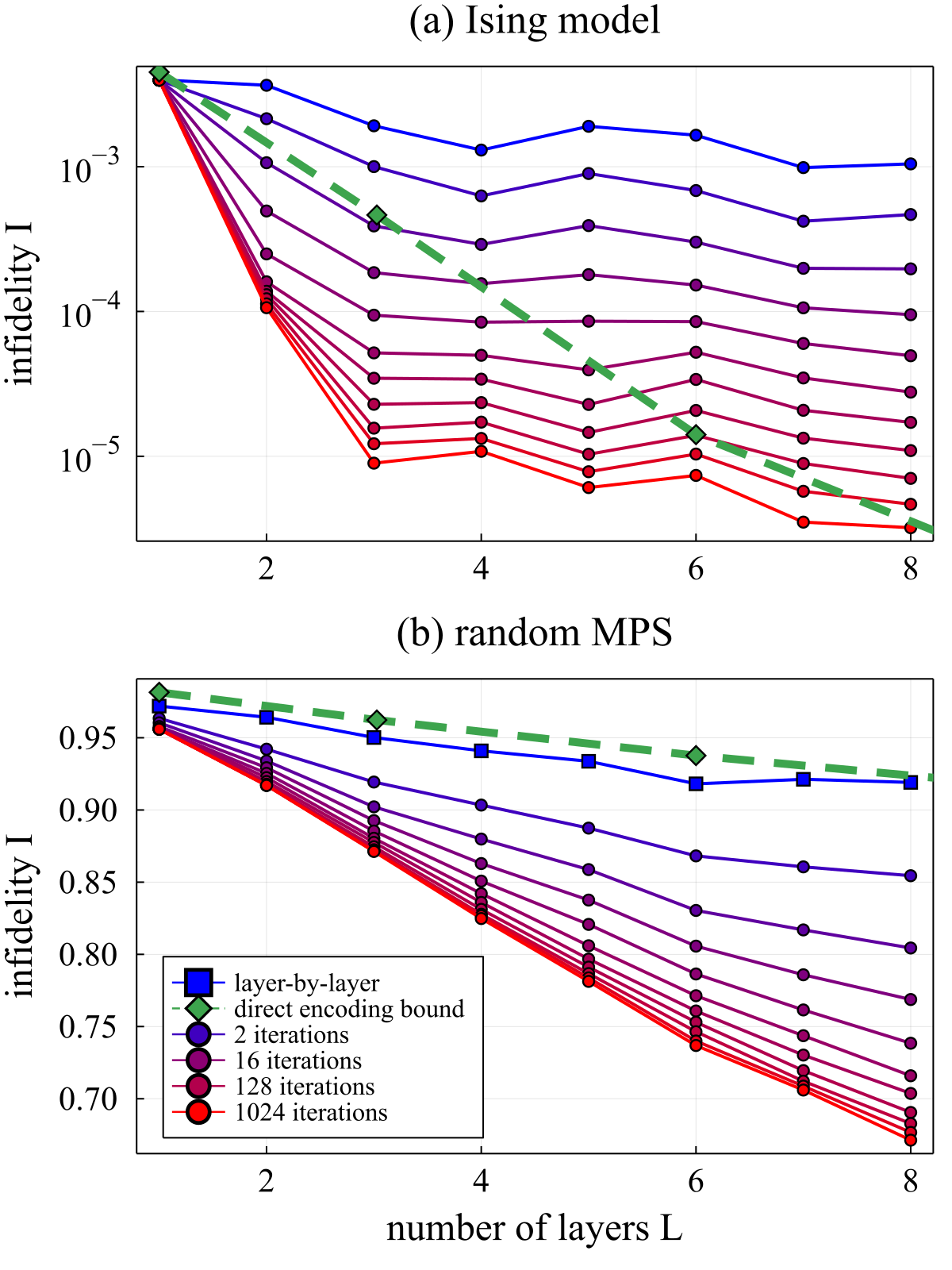}
    \caption{Infidelity $I$ as a function of the number of layers $L$ for two target states with $N=12$ qubits and bond dimension of $\chi=64$: (a) low entanglement ground state of the Ising model; (b) random MPS state. The obtained infidelity for different number of iterations is compared to the layer-by-layer approach \cite{Ran_2020}, and to the direct bound given in the main text. }
    \label{fig:infidelity}
\end{figure}

A more severe problem is that the layer-by-layer approach is not effective: after the first few layers, any additional layer gives a negligible improvement.
To highlight this point, we probe the infidelity $I=1-F$, where the fidelity $F=|f|^2$, and $f$ is the overlap of the encoded state with the target state $\ket{\psi_{\rm target}}$,
\begin{align}
	f(\lbrace G^l_i\rbrace) = \langle{\psi_{\rm target}}|{\prod_{l=1}^{L} \prod_{i=1}^{N}}G_i^l|{0}\rangle.\label{eq:overlap}
\end{align}
Fig.~\ref{fig:infidelity} shows in blue the infidelity of the layer-by-layer approach as a function of the number of layers, for two target states with $N=12$ qubits and bond dimension $\chi = 64$ \footnote{See also \ref{sec:additional} for additional results using other system sizes, showing consistent results}: (a) the ground state of an Ising model \footnote{The Hamiltonian of the Ising model is $H = \protect\sum_{n}{J_z S^z_n S^{z}_{n+1} - h_x S^x_n}$. This Hamiltonian is gapped for all $h_x\neq \pm J_z$. Its ground state is an example of a non-trivial state with low entanglement, which is faithfully described by MPS. Here, we consider a ferromagnetic state with $h_x/J_z=0.6$}; (b) a random MPS. In both cases, the layer-by-layer approach shows a very slow convergence to the target state, as a function of the number of layers $L$. In fact, it is not even clear that the layer-by-layer algorithm will eventually converge to $I\to 0$ as $L\to\infty$. Moreover, we observe that the layer-by-layer approach is not capable of reproducing the correct circuit, even in cases where the target state is created using a small number of layers only. This is due to the random kernel used in the algorithm and to its inability to ``plan ahead'' and create cooperation between different layers. In each step, the algorithm maximizes only the immediate overlap of the state, without taking other amplitudes into account, which may potentially facilitate the performance of future (deeper) layers.

\section{Optimization algorithm for encoding MPS states\label{sec:classi_opt}}

To improve the layer-by-layer approach, we introduce an optimization protocol that aims at finding the best circuit for a fixed number of layers and a fixed circuit topology. Developing a deterministic optimization algorithm requires us to evaluate the gradient of the cost function with respect to the variational parameters, $G_i^l$. Fortunately, the gradient of the overlap $f$ is simply given by the original tensor network with one missing tensor,
as illustrated in Fig.~\ref{fig:tens_cost_func}(b). The resulting circuit can, therefore, be efficiently contracted into a single rank-4 tensor with $2^4=16$ elements, using tensor-network methods \footnote{The gradient described above is the unconstrained gradient, which does not take into account the unitarity requirement of the gate $G_i^l$. Gradient descent optimization algorithms follow only the part of the gradient which preserves unitarity, as detailed in \ref{app:gradient}, see also further reviews in Refs.~ \cite{manton2002optimization,boumal2022intromanifolds, absil2009optimization}}.
%
We used this gradient to perform an element-by-element optimization scheme. This method, which has been used at first for optimization of sequentially generated entangled multi-qubit states and MPOs in Refs.~\cite{saberi2009constrained, saberi2011ancilla}, is known also as the Evenbly-Vidal method for tensor network renormalization \cite{evenbly2015tensor}. It has been since used in the context of variational quantum algorithms \cite{lubasch2020variational}, quantum simulations \cite{Lin_2021}, and tensor network optimization  \cite{hauru2021riemannian}. The method is  analogous to the coordinate-descent method in classical machine learning \cite{wright2015coordinate} and to the iterative update of tensors in the DMRG algorithm \cite{schollwock2011density} in the effort to optimize the objective function by parts. Unlike steepest descent methods, the element-by-element optimization does not rely on a convergence rate which needs continuous adjustment using heuristics, but rather proceeds through well defined and discrete steps. At each step, we select a single gate, $G^l_i$, and find the unitary matrix that maximizes ${\rm Tr}[(\partial f/\partial G^l_i)G^l_i]$. This matrix can be obtained by computing the tensor $\partial f/\partial G_i$ and  projecting it to the closest unitary matrix. This last step simply involves the calculation of the SVD decomposition of the gradient and the substitution of the SVD values by 1. We, then, substitute the gate with the new optimum and proceed to the next gate. When all gates have been updated, the algorithm moves to the next iteration and runs again the protocol over all gates.  The algorithm is repeated several times, until a specific convergence criterion is reached.

As shown in Fig. \ref{fig:infidelity}, the proposed optimization protocol leads to a significant improvement over the layer-by-layer approach: (a) For the ground state of the Ising model, we achieve an improvement of up to one order of magnitude with just $10$ iterations. We observe that the improvement is roughly inversely proportional to the number of iterations (the curves are equally spaced on a logarithmic scale). Hence, our protocol is efficient, in the sense that it can deliver infidelities arbitrarily close to the optimal value in a polynomial time. (b) For random states, the variational circuit is not expressive enough to achieve low values of the infidelity. 
Nevertheless, we observe that our optimization algorithm achieves a significant improvement over the layer-by-layer scheme and converges after a small number of iterations. 

To check the efficiency of our method, we compare it to a more direct approach consisting of, first, reducing the size of the MPS tensors to a truncated bond dimension $\chi_0$, and then, encoding them state using n-qubit gates. As mentioned above, a MPS of bond dimension $\chi_0$, can be directly converted into a multi-qubit gate acting on $k = {\rm log}_2 \chi_0 + 1$ qubits. Constructing a general k-qubit gate requires using many 2-qubit gates, whose number can be bounded from below by the formula $\Omega(k) =  \frac{1}{9} 4^k - \frac{1}{3} k - \frac{1}{9} $ \cite{barenco1995elementary}, which is derived by counting the number of degrees of freedom encoded in a single k-qubit gate. Combining the lower bound and the the relation between $k$ and $\chi_0$, we obtain that using the direct approach, with truncated bond dimension $\chi_0$ requires the same number of 2-qubit gates as the optimized circuit with $L = \frac{4}{9}\chi_0^2  -\frac{1}{3}{\rm log}_2(\chi_0)-\frac{4}{9}$ layers \footnote{The comparison ignores the reduced number of gates at the edges of the system, which can lead to a minor difference in the total number of gates}. In particular, the case of $\chi_0=2$ is equivalent to a single layer of 2-qubit gates, $L=1$, while $\chi_0=4$ corresponds to 3-qubit gates and requires number of layers of $L=6$. Although this formula is originally valid for full k-qubit gates, which are equivalent to $\chi = 2^{k-1}$,  it can also be viewed as a lower bound for any value of bond dimension. For an integer $\chi$ the number of degrees of freedom is proportional to $\chi^2$, and the number of 2-qubit gates required to encode  them can be counted similarly to k-gate encoding.
 
In Fig~\ref{fig:infidelity} we compare the infidelities obtained the direct method (dashed line) with  those obtained by our optimization method (continuous lines). For the Ising model ground state, we find that at low depth the optimized infidelity is significantly lower than the infidelity of the direct encoding, while at higher depths the direct encoding is nearly optimal.
In contrast, for the random MPS, the direct truncation method performs poorly in comparison to the optimized circuits. This comparison highlights that the optimized circuits takes into account higher singular values of the MPS which the truncation omits, and when the tail of eigenvalues is significant, they perform much better than the direct encoding approach.

\begin{figure}
\centering
\includegraphics[width=1.1\columnwidth]{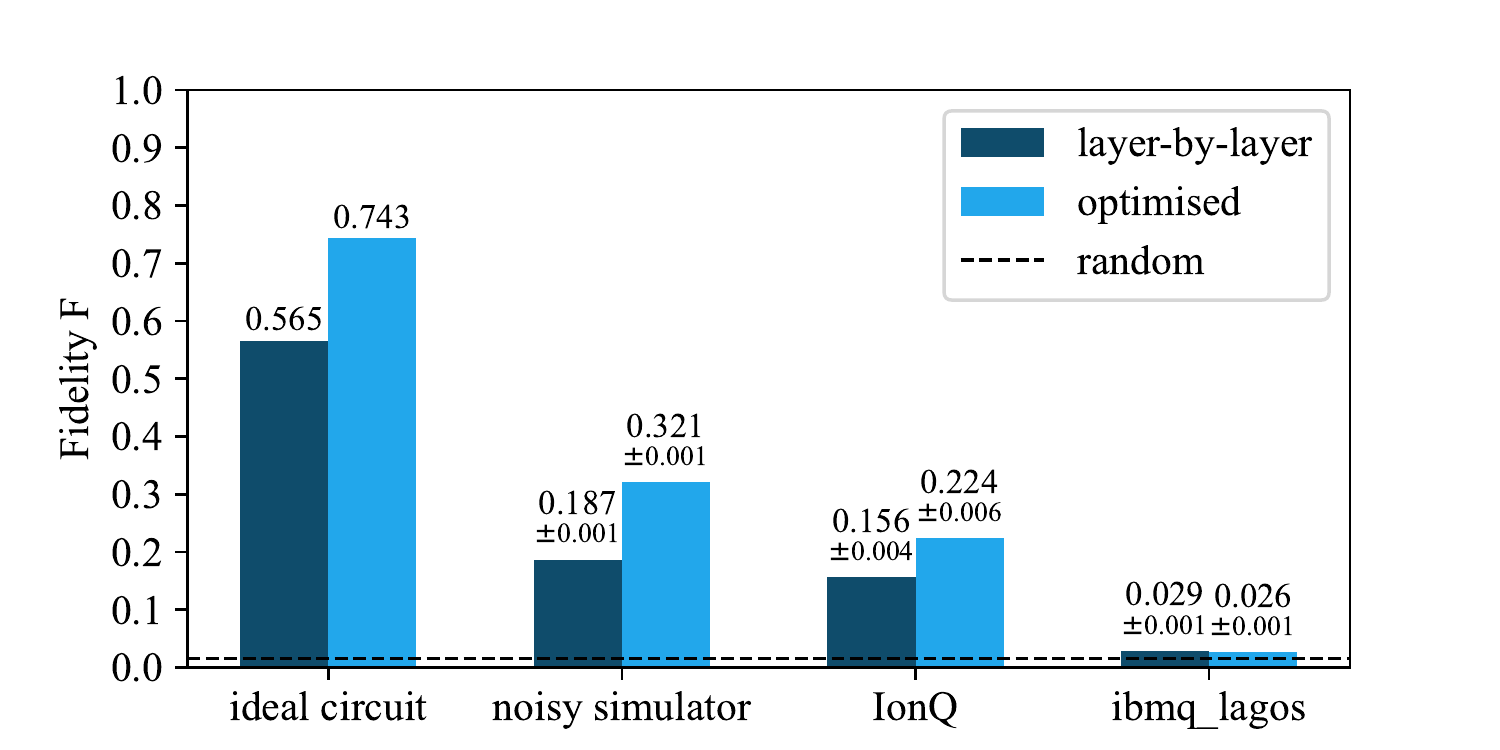}
\caption{Fidelity of encoding a random-MPS target state of $N=6$ qubits using circuits with $L=3$ layers, obtained by the layer-by-layer method of Ref.~\cite{Ran_2020} and by the present optimization algorithm.}
\label{fig:experiment}
\end{figure}

Our optimization algorithm works under the assumption of ideal 2-qubit gates and an important question is what happens in a real noisy device. Due to the limited capabilities of available quantum processors, we considered the minimal instance of a target state where the proposed optimized circuit is expected to achieve a significant advantage over the layer-by-layer method, namely a random-MPS state with $N=6$ qubits, encoded with $L=3$ layers, see Methods section for details. Fig.~\ref{fig:experiment} compares the ideal fidelity with the results obtained in a noisy simulator and on quantum computers by IonQ and IBM.
 The circuit obtained by our optimization method outperformed the layer-by-layer encoding, except in the case of the IBM Lagos quantum processor, whose faulty result is comparable to the fidelity of a random state, $F= 1/2^{N}$. As discussed earlier, the difference between the two encoding strategies is expected to grow with the accessible circuit depth. 

\section{State encoding on a quantum computer \label{ch:quantum-opt}}



Up to this point, we have assumed that the target state is known classically and all matrix elements can be computed using efficient tensor networks. 
We now move to a situation where the target state is not known classically, but can be prepared on a quantum computer. Encoding a state directly on a quantum computer is a desired technique for several reasons: For example, this scenario is common in the study of nonequilibrium quantum effects, such as quantum quenches where one aims at studying the time evolution of an initial state \cite{calabrese2006time,polkovnikov2011colloquium,mitra2018quantum,paeckel2019time,Lin_2021}. In addition, this approach is useful for the compilation of the initial layers of quantum machine learning algorithms \cite{schuld2018information,leymann2020bitter,Endo_2021, cerezo2021variational}. Finally, it enables one to store the output of complex quantum calculations by obtaining a faithful shallow-circuit representation \cite{liu2020variational,dilip2022data}.

\begin{figure}[t]
    \centering
    \includegraphics[width=\columnwidth]{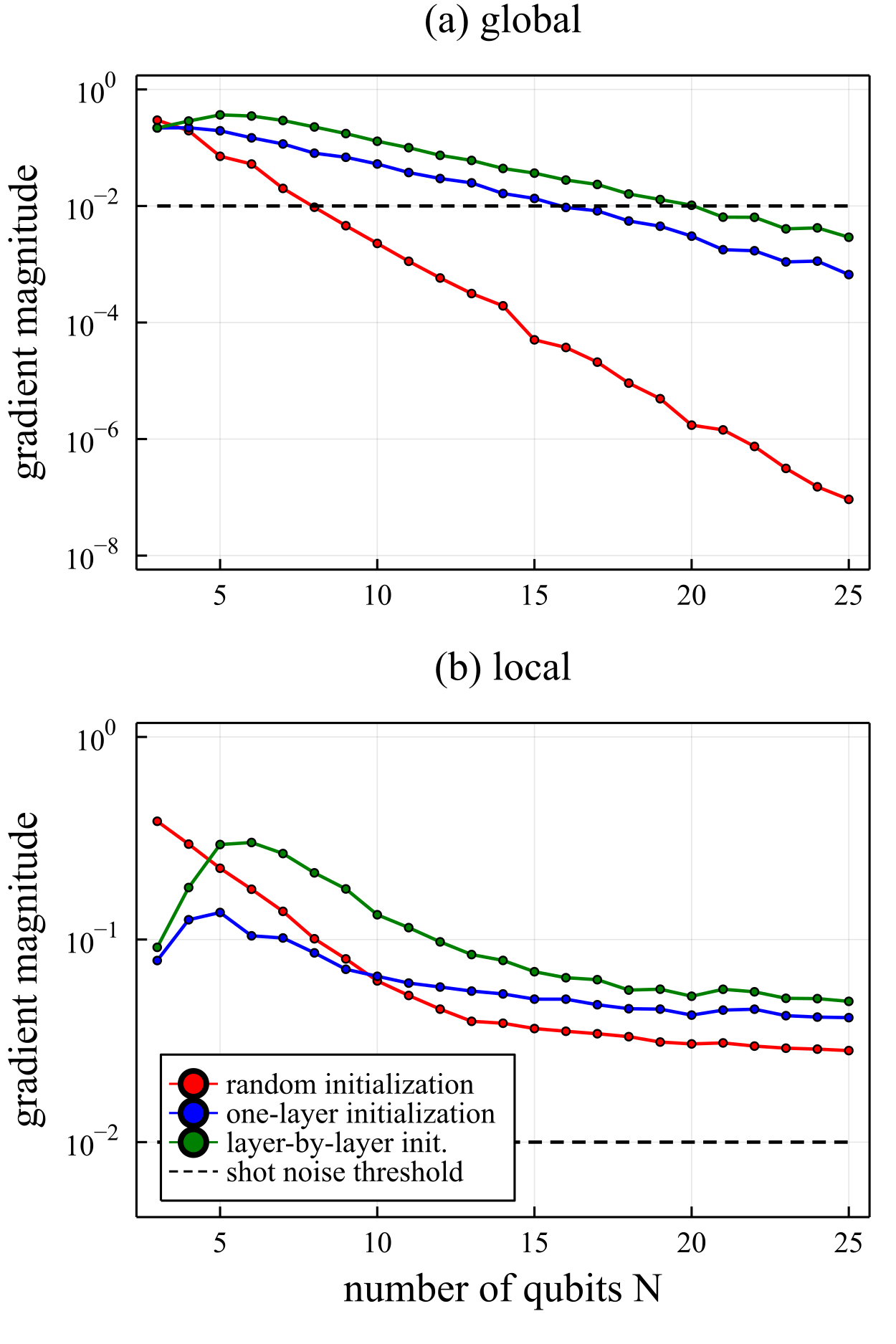}
    \caption{Magnitude of the unitary part of the initial gradient  $\| {\rm grad}_{G}F \|$ for $L = 5$ layers, as a function of the number of qubits, $N$, for different initalizations: (a) global cost function and (b) local cost function. In the global cost function (a), the gradient decreases exponentially with $N$ for all initializations, while in the case of a local cost function, it is polynomial. The dashed line represents the threshold for resilience to realistic shot noise (see text).}
    \label{fig:grad_5_layers}
\end{figure}


In all these applications, the goal is to encode the target state using a smaller number of gates \footnote{Here we assume that a preexisting faithful copy of the desired state has already been encoded in the quantum computer. Otherwise, training the ansatz on a quantum computer might not be scalable, as a cost function that is based on comparisons to tomography measurements in multiple bases will likely result in barren plateaus, due to its global nature.}.
In a real quantum processor, the tensor derivative of the circuit cannot be performed analytically, but needs to be evaluated using tensor tomography. This process, which we refer to as gradient tomography, can be performed by measuring all possible products of Pauli matrices acting on the four sites in red in Fig.~\ref{fig:tens_cost_func}(b)~\footnote{See \ref{app:tomography} for more details}. Due to the finite number of shots used to perform the tomography, the result of this procedure will include random fluctuations. For a typical number of shots of the order of a few thousands, the resulting uncertainties are of the order of few percents. The optimization algorithm is expected to work only as long as the gradients are large enough for the first steps of the optimization process. 

Under generic conditions, the overlap between the initial configuration of the variational circuit and the target state decreases exponentially with the number of qubits. As a consequence, the amplitude of the gradient is exponentially suppressed and may not be detected experimentally. This situation, often referred to as barren plateaus \cite{mcclean_barren_2018, cerezo2021cost,holmes2022connecting}, is exemplified in Fig.~\ref{fig:grad_5_layers}(a) for the case of random MPS target states with bond dimension $\chi = 64$, encoded using $L=5$ layers. The figure displays the magnitude of the unitary gradient after the initialization of the circuit ansatz, at the beginning of the optimization algorithm. In the case of a random initialization, the gradient amplitude vanishes exponentially with the number of qubits, and follows the same slope as the average fidelity between random states, $F\sim 1/2^{N}$. 

If the state is known in its MPS representation, one can try to beat the barren plateau by using a better initialization of the unitary gates. A smart initialization may bring the system to the proximity of the global minimum of the cost function, where the gradients are more pronounced. In our case, we can obtain a better initialization using the layer-by-layer algorithm.  Fig.~\ref{fig:grad_5_layers}(a) shows the unitary gradient obtained by applying this algorithm to a single layer (blue) or to all 5 (green). Due to the initialization the gradient is greatly increased, but still vanishes exponentially $\sim 0.75^{-N}$. For $N>15$ the gradient already reaches the threshold value of $10^{-2}$ (dashed line), beyond which shot noise becomes dominant and the optimization protocol is stuck in a barren plateau.


A common technique to address barren plateaus 
is to substitute the {\it global} cost function, $I$, with a {\it local} one. This approach was shown mathematically to solve the barren plateaus problem for shallow quantum circuit, i.e.\ for circuits where the circuit depth does not grow with the number of qubits  \cite{cerezo2021cost}.  Note that the mathematical theorems assumed a brick-wall gate structure, which is different than the present staircase gate connectivity. According to the definition of Ref.~\cite{cerezo2021cost}, the present circuit has depth $L+N$ and, hence, beyond the regime of validity of the rigorous theorem. As we now show, our numerical calculations give evidence that the local cost function approach is valid in this case as well.
We implement this strategy by introducing the local infidelity, $I_{\rm local}=1-\sum_{n=1}^N F_{\rm local,n}/N$, with
\begin{align}
F_{\rm local,n} = \bra{\psi_{\rm target}} \prod_{i,l} G_i^l P_n \prod_{i,l} \left(G^l_i\right)^\dagger \ket{\psi_{\rm target}}\label{eq:local}
\end{align}
as illustrated in Fig~\ref{fig:tens_cost_func}(c), where $P_n = \ket{0}_n\bra{0}_n$ is the projection over the $0$ state of the $n$th qubit. The fidelity $F_{\rm local, n}$ represents the probability to measure the $n$'th qubit in the 0 state, irrespective of the state of the other qubits, and is maximal (equals to 1) when the circuit prepares the target MPS exactly.
As shown in Fig.~\ref{fig:grad_5_layers}(b), the usage of a local cost function solves the barren plateaus  for our problem: the gradient is no longer an  exponential function of the number of qubits and does not cross the shot-noise threshold. 

\begin{figure}
    \centering
    \includegraphics[width=\columnwidth]{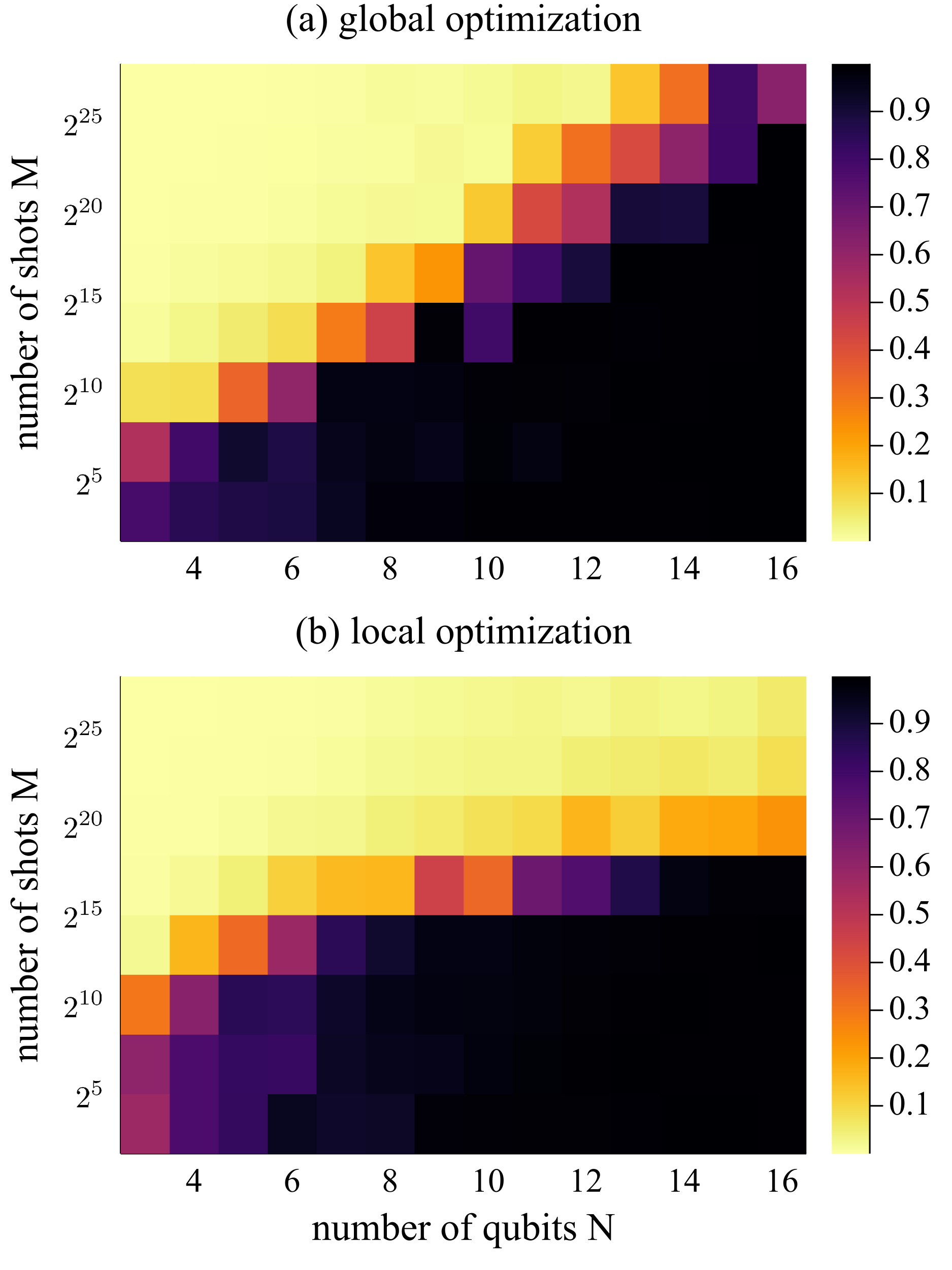}
    \caption{Infidelity $I$ obtained by 20 iterations of the global and local optimization algorithms, as a function of the number of qubits and shots, for a random-MPS target state with $L=5$ layers and $\chi=64$.}
    \label{fig:shot_noise_map}
\end{figure}

A potential drawback of the local cost function is that one needs to compute the derivative of quadratic functions $F_{\rm local,n}$, rather than a linear one, $f$, see Eqs.~(\ref{eq:overlap}) and (\ref{eq:local}). As a consequence, finding the optimal gate requires the measurement of a larger tensor, potentially inducing more errors in the calculation. To estimate this effect, we test our optimization protocol on a phenomenological model that takes into account the finite number of shots per measurement, which we denote by $M$, see Methods section for details. Fig.~\ref{fig:shot_noise_map} compares the fidelity obtained using the global and local cost functions, for random-MPS target states. We find that, in the case of a small number of qubits $N\leq 8$, the local optimization is penalized by the larger number of measurements and leads to a larger infidelity. In contrast, for a large number of qubits, the local method performs better, as the polynomial scaling of the local gradient wins against the exponentially decreasing global gradient. The color coding (bright for small infidelities and dark for large infidelities) enables for a clear separation between successful and unsuccessful instances: for the global cost function one needs an exponential number of shots, while for the local one, a polynomial number is sufficient.

\section{Conclusion and discussion}

In this work, we addressed the problem of quantum state encoding, namely how to encode a target state using 1-qubit and 2-qubit gates only. We opened our study with MPSs, an efficient representation of quantum states with low levels of entanglement. Realizing these states on a quantum computer requires a circuit depth that grows at least linearly with the bond dimension $\chi$.  
Here, we showed how to create a state that approximates the MPS using a smaller number of gates.
Our approach focuses on quantum circuits with a fixed depth and aims at finding the gate configuration that offers the best approximation to the MPS. We described an optimization technique that iteratively improves each gate, until the optimal solution is found. Each step involves a singular value decomposition of a small tensor, whose elements can be computed efficiently using tensor network techniques and require limited resources only. By considering two extreme types of MPSs (low-entanglement states and random states), we demonstrated that our algorithm converges to a solution with a good fidelity. Empirically, we observed that the algorithm is not trapped in local minimum, unlike other variational approaches where local minima are abundant \cite{meister2022exploring}.

Next, we considered the realization of the proposed element-by-element optimization algorithm directly on a quantum computer. This task is relevant to the situations where the MPS is not known classically, such as in the case of a quantum dynamics simulations, encoding the outputs of complex quantum calculations. In a quantum computer the tensors can be estimated up to a finite resolution only, due to the finite number of measurements. In this case, the optimization algorithm encounters barren plateaus. We explained how to solve this problem by moving to a local cost function, and demonstrated its performance by considering the effects of realistic shot noise.

The main outcome of our work is a generic algorithm that allows one to encode a quantum state using a small number of gates. Unlike gradient-based optimization algorithms, this algorithm is monotonic, in the sense that each step necessarily brings us closer to the final goal. Its key ingredient is the direct estimation of the tensor derivative of the circuit and its translation into an improved set of gates. An important question that deserves future investigation is how to  choose the circuit depth and the gate connectivity that will deliver the largest fidelity. On the one hand, by considering a larger number of layers and a more complex connectivity, one expects to obtain  a larger fidelity. On the other hand, in the real world, adding more gates leads to more decoherence and noise, and, hence, to a smaller fidelity. Finding the optimal number will generically depend on the entanglement of the target state and on the gate faultiness of the device. Studying this dependence theoretically and experimentally is a fundamental step towards demonstrating the applicability of our approach.



\section{Methods}
{\bf Implementation on a real device}
To test our method for MPS encoding on a real quantum processors we chose a quantum state for which the difference between the layer-by-layer encoder and the optimized encoder is significant. Due to the limited capability of state-of-the-art devices, we chose a small random state with $N=6$ qubit and encoded it using $L=3$ layers \footnote{For smaller systems, the maximal bond dimension is $\chi=4$ and the difference between one and two layers is very small, such that the benefit of our optimization is not significant}. We sampled 1000 random states and calculated the fidelity of the layer-by-layer encoder for each one. We, then, chose the random state that exhibited the lowest fidelity and applied our optimization algorithm to obtain a better encoder. The encoder circuits were transpiled into a sequence of native gates of single qubit rotations and CNOT gates using Qiskit libraries, which implement an exact 2-qubit gate decomposition by a sequence of 3- CNOT gates interleaved with 1-qubit rotation gates, as detailed in \cite{Zhang_2003}. The original (optimized) transpiled circuit contained 142 (163) 1-qubit rotations, and 23 (24) CNOT gates.
Next, we performed full quantum state tomography of the output state, using $3^6=729$ circuits, with at least 200 measurement shots each. From these measurements, we estimated the density matrix of the final state, $\rho_{\rm exp}$, and computed the fidelity to the target state as $F=\langle \psi_{\rm MPS}|\rho_{\rm exp}|\psi_{\rm MPS}\rangle$.

The noisy simulation was performed using QISKIT Aer with noise parameters extracted from the ibm\_lagos quantum processor. As shown in Fig.~\ref{fig:experiment}, the actual device performed much worse and we attribute this discrepancy to the limited connectivity of the device: our algorithm assumes a linear connectivity between the 6 qubits, which is unavailable on the device. The QISKIT compiler solves this problem by adding SWAP gates between qubits that are not directly connected, but these extra gates lead to a much worse performance.

{\bf Realistic model of shot noise} Our optimization algorithm is based on the measurement of local gradients, obtained by disconnecting a single gate from the circuit. 
To estimate the effects of shot noise, we expanded the gradient tensors in sums of Pauli strings and, then, replaced the coefficient of each Pauli string by a random variable sampled from a binomial distribution with $M$ trials. This substitution left  the average value of the tensor unchanged and mimicked the result of a quantum measurement with $M$ shots. Our model disregards the effects of phase noise, which may lead to a further suppression of the gradient, but is sufficient to obtain a qualitative estimate of the effects of shot noise on the gradients.

\section{DATA AVAILABILITY}
All data and figures that support the findings of this paper are available
from the corresponding author on request. Please refer to Matan Ben-Dov at matan.ben-dov@biu.ac.il.

\bibliographystyle{naturemag}
\bibliography{main}

\begin{acknowledgments}
We thank Yossi Avron, Itai Arad, Fabrice Frachon, Haggai Landa, Frank Pollman, Matt Reagor, Efrat Shimshoni and the SQMS alogrithm team for useful discussions. The IonQ quantum computer was accessed through the Microsoft Azure Quantum online service and we acknowledge their financial support. This research is supported by the Israel Science Foundation, grants number 151/19 and 154/19. 
\end{acknowledgments}

\section{Author Contributions}
E.D.T. and A.M. conceived the idea for this work. E.D.T. and M.B.D. developed the theoretical framework detailed in this work, M.B.D designed and performed the simulations and analyzed their results.  Diagrams were created by E.D.T and figures were created by M.B.D. D.S performed and analyzed the experiments on quantum computers.M.B.D and E.D.T wrote the manuscript, and A.M revised it.

\section{COMPETING INTERESTS}
The authors declare no competing financial or non-financial interests.

\newpage

\renewcommand{\theequation}{S\arabic{equation}} 
\setcounter{equation}{0}

\renewcommand{\thefigure}{S\arabic{figure}}   
\setcounter{figure}{0}
\renewcommand{\thesection}{SI-\arabic{section}}   
\setcounter{section}{0}
\renewcommand{\thetable}{S\arabic{table}}   
\setcounter{table}{0}

\onecolumngrid

\newpage

\begin{center}
\large{\bf Supplementary Information\\ for ``Approximate encoding of quantum states using shallow circuits'' }
\end{center}
\section{Gradient with respect to a unitary matrix}
\label{app:gradient}

In the main text we described an optimizations algorithm using unitary matrices as variables, without the need to use an explicit parameterization of the different gates. This approach gives us an insight into the connection between gates connectivity, quantum information, tensor-networks and the performance of the circuit as a quantum state encoder. In this section, we provide more information on the definition of the derivative according to a general complex matrix and how it compare to the constrained derivative, the derivative in respect to a matrix compliant with a unitarity constraints, see also Refs.~  \cite{manton2002optimization,boumal2022intromanifolds, absil2009optimization}. 

First we will define the directional derivative of a function $f(\{G_n\})$ according to the coordinate $G_i$. The directional derivative measures the change in the function when $G_i$ is infinitesimally modified in a specific direction. The directional derivative evaluated at the point ${G_n}$ in the direction of $\Delta$ is
\begin{equation}
    \frac{\partial f}{\partial G_i} |_{\{G_n\}}(\Delta) = \lim_{\epsilon\to 0}{\frac{f(\{G_n+\epsilon \delta_{n,i}\Delta\}) - f(\{G_n\})}{\epsilon}}.
\end{equation}

This is a partial derivative as we only varied the $G_i$ coordinate. 

To define the gradient, we need to consider how the directional derivative along an arbitrary direction can be derived  as a projection of a single object on the direction of the derivative. This link is expressed through an inner product 
\begin{equation}
\label{eq:direc_grad_connection}
     \frac{\partial f}{\partial G_i} |_{\{G_n\}}(\Delta) = \langle  \nabla_{G_i} f,\; \Delta \rangle.
\end{equation}

In our case, we consider real functions over the domain of complex matrices, whose inner product is the real part of Hilbert-Schmidt inner product $\langle A,
B\rangle = \re{\mathrm{Tr}(A^\dagger B)}$. Taking the real part of the trace will be essential to computing the gradient for different function, as we will see below.
To calculate the gradient, one needs to calculate the directional derivative for general direction, compare it with the inner product and extract gradient matrix from eq. \ref{eq:direc_grad_connection}.

We now describe a few relevant examples for our cases.
The first example is a function over the real matrices- $f_{\rm overlap}(A) = \mathrm{Tr}(E^\intercal A)$ is a simple linear function in respect to a real matrix $A$. Its directional derivative is
\begin{align*}
    \frac{d f_{\rm overlap}}{d A}(\Delta) &= \lim_{\epsilon \to 0}{\frac{\mathrm{Tr}(E^\intercal(A+\epsilon \Delta)) - \mathrm{Tr}(E^\intercal A)}{\epsilon}} \\
    &= \frac{\epsilon \mathrm{Tr}(E^\intercal\Delta)}{\epsilon}\\
     &= \mathrm{Tr}(E^\intercal\Delta),
\end{align*}
and comparing the result to eq. \ref{eq:direc_grad_connection} (now using the real version of the inner-product), we can conclude that in this case $\nabla_{A} f_{\rm overlap} = E$, as expected from a linear function.


For a function over the complex matrices, perfect linearity is not possible, but it is instructive to examine a similar function which takes the absolute value of the function from before- $f_{\rm \left|overlap\right|}(G) =\left| \mathrm{Tr}(T^\dagger G)\right|$, when now $G$ and $T$ are both complex matrices.
In this case,
\begin{align*}
    &\frac{d f_{\rm \left|overlap\right|}}{d G}(\Delta) = \lim_{\epsilon \to 0}{\frac{\left|\mathrm{Tr}(T^\dagger(G+\epsilon \Delta))\right| - \left|\mathrm{Tr}(T^\dagger G)\right|}{\epsilon}} \\
    &= \lim_{\epsilon \to 0}{\frac{\sqrt{\left|\mathrm{Tr}(T^\dagger G)\right|^2 + 2\epsilon \re{{\mathrm{Tr}(T^\dagger G)}^*\mathrm{Tr}(T^\dagger \Delta)}+\epsilon^2 \left|\mathrm{Tr}(T^\dagger \Delta)\right|^2} - \left|\mathrm{Tr}(T^\dagger G)\right|}{\epsilon}} \\
    &= \lim_{\epsilon \to 0}{\frac{\left|\mathrm{Tr}(T^\dagger G)\right|\left(1 + \frac{1}{2}2\epsilon \frac{\re{{\mathrm{Tr}(T^\dagger G)}^*\mathrm{Tr}(T^\dagger \Delta)}}{\left|\mathrm{Tr}(T^\dagger G)\right|^2}\right) - \left|\mathrm{Tr}(T^\dagger G)\right|}{\epsilon}} \\
    &= \frac{\re{{\mathrm{Tr}(T^\dagger G)}^*\mathrm{Tr}(T^\dagger \Delta)}}{\left|\mathrm{Tr}(T^\dagger G)\right|} \\
    &= \re{\mathrm{Tr}\left(\frac{{\mathrm{Tr}(T^\dagger G)}^*}{\left|\mathrm{Tr}(T^\dagger G)\right|} T^\dagger\Delta \right)},
\end{align*}
and at the end the extracted gradient is $\nabla_{G} f_{\rm \left|overlap\right|} = \frac{\mathrm{Tr}(T^\dagger G)}{\left|\mathrm{Tr}(T^\dagger G)\right|}T$, which is the same as the linear function shown in the previous example up to a complex phase.

Another important example is the quadratic function $f_{\rm infidelity}(G) = 1- \left|\mathrm{Tr}(T^\dagger G)\right|^2$, which gives the directional derivative of
\begin{align*}
    &\frac{d f_{\rm infidelity}}{dG}(\Delta) =\\
    &=\lim_{\epsilon \to 0}{\frac{-\mathrm{Tr}(T^\dagger(G+\epsilon \Delta))\mathrm{Tr}((G+\epsilon \Delta)^\dagger T) + \mathrm{Tr}(T^\dagger G)\mathrm{Tr}(G^\dagger T)}{\epsilon}} \\
    &=\lim_{\epsilon \to 0}{\frac{-\epsilon \mathrm{Tr}(T^\dagger \Delta) \mathrm{Tr}(G^\dagger T)-\epsilon \mathrm{Tr}(T^\dagger G) \mathrm{Tr}(\Delta^\dagger T) + o(\epsilon^2)}{\epsilon}} \\
    &=-\mathrm{Tr}(T^\dagger \Delta) \mathrm{Tr}(G^\dagger T) - \mathrm{Tr}(T^\dagger G) \mathrm{Tr}(\Delta^\dagger T)\\
    &=-2\re{\mathrm{Tr}(G^\dagger T)\mathrm{Tr}(T^\dagger \Delta)},
\end{align*}
and comparing the derivative to gradient definition we get $\nabla_{G} f_{\rm overlap} = -2\mathrm{Tr}(T^\dagger G)T$.

Next, we consider the constrained case of unitary matrices. In this case, the above-mentioned gradient will not deliver the direction of the steepest slope, as stepping in the direction of the {\bf free} gradient does not conserve the unitarity of the matrix.  The unconstrained gradient should be modified to conserve the unitarity property. This can be obtained by projecting the gradient onto the tangent space of the unitary manifold $U(n)$ from which the matrices are taken
\begin{equation}
    \mathrm{grad} f = {\rm proj}_{G}(\nabla(f)).
\end{equation}
The projection can be viewed as filtering the gradient to keep only the part that conserve unitarity condition:
\begin{equation}
    G^\dagger G \to (G + \epsilon\Delta)^\dagger(G + \epsilon\Delta) = I
\end{equation}
For the conservation to hold up to the first order in $\epsilon$, the resulting condition is $G^\dagger\Delta = -(G^\dagger\Delta)^\dagger$, or equivalently, demanding that $G^\dagger\Delta$ is anti-hermitian. A projection of a matrix $U$ onto the tangent space ${T}_{G}{M}$ is therefore
\begin{equation}
    \mathrm{proj}_{G}(U) = G\frac{1}{2}(G^\dagger U - U^\dagger G) = \frac{1}{2}(U - G U^\dagger G),
\end{equation}

and the formula of the constrained gradient for unitary matrices is 
\begin{equation}
\label{eq:unitary_grad}
    \mathrm{grad} f = \frac{1}{2}\left(\nabla f  - G (\nabla f)^\dagger G\right).
\end{equation}

\section{The Steepest Descent and Element by element optimization algorithms}
\label{app:element}

In the main text we mentioned two optimization methods for our general 2-qubit gate ansatz: steepest descent and element-by-element optimization. We will now expand on the two methods and their implementation for optimization of global and local cost functions.

In the gradient descent algorithm the full gradient is calculated with respect to the different unitary gates, after which the gates are updated by a small step in the direction of the gradient. Usually the gradient is subtracted from the coordinate vector according to 
\begin{equation}
	G^l_i\rightarrow G^l_i - \eta~\mathrm{grad}_{G_i} F |_G,
\end{equation}
where \(\eta\) is the step size which often varies adaptively along the optimization process, and $\nabla_i F$ is the unitary gradient along $G_i$ as defined in eq. \ref{eq:unitary_grad}.

For small step sizes $\eta \ll 1$ the descent in the direction of the gradient conserves unitarity up to first order in $\eta$, and the algorithm can be performed by infinitesimal steps, according to
\begin{align}
     G_i \to G_i - \epsilon\left(\nabla_i F -  G_i\nabla_i F^{\dagger}{G_i}\right),
\end{align}
which results in a "gradient flow" optimization, similar to the method described in \cite{wiersema2022optimizing}. 

For large steps size, it is necessary to enforce unitarity after each step of optimization by projecting the new gate back to the unitary manifold $U(n)$ using
\begin{equation}
	G^l_i\rightarrow \left(G^l_i -  \eta~\mathrm{grad}_{G_i} F \right)_{\rm unitary}.
\end{equation}
The projection brings the matrix to the closest unitary counterpart according to
\begin{equation}
\label{eq:SVD}
	(T)_{\rm unitary}=(USV^\dagger)_{\rm unitary} = UV^\dagger.
\end{equation}
Here, we used the singular value decomposition (SVD) of the matrix and substituted all the singular values with 1, according to the unitarity requirement. It is possible to verify this formula by checking that it offers a solution of the minimization of the trace distance between the matrix $T$ and any unitary matrix $U$ (as seen in the examples in \ref{app:gradient}), by checking that the unitary gradient is zero for this particular choice.
Another possiblity is performing the optimization by moving a step on a geodesic in the direction of the gradient, which is outside the scope of this paper.


There are several differences between gradient descent performed on 
unitary gates and gradient descent on angle parameterization of the gates. Firstly, using the projection of a linear step on the unitary manifold might cause limited reachability for a single optimization step, and thus the number of steps needed to get to the proximity of the minima can be larger if the initial guess is far away. 
On the other hand, using gate parameterization induces directional biases to optimization process. For further details the reader may refer to \cite{birtea2022constraint, wiersema2022optimizing}.

The element by element optimization is based on the notion of breaking apart the bigger optimization problem to small tasks that can be solved analytically. By optimizing a single gate $G^l_i$ each time while fixing the others, we can find the single gate the yields the lowest cost with respect to the selected cost function. This sub-optimization can be calculated analytically when the cost function is linear or quadratic in $G_l^i$, which can speed up the optimization process considerably.

The gradient descent algorithm is agnostic to the type of the cost function and can be applied to both the global and local cost function described in the paper.
In contrast, the element-by-element optimization algorithm has to be adjusted to according to the choice of the cost-function.
For the case of global cost function, we take the cost function to be the infidelity between the encoded state and the target state:
\begin{equation}
    F_{\rm global}(\{G^l_i\}) = 1 -  \left|f(\{G^l_i\})\right|^2.
\end{equation}
In this case we can choose to optimize the overlap function $\left|\mathrm{Tr}(G^l_i\nabla^i_l F)\right|$ and find its maximum, which will coincide with the minimum value of the cost function. Looking at the overlap is useful here, as up to a complex phase the overlap contains only linear terms in $G_i^l$ which simplify the optimization algorithm.
The overlap function can be expressed in terms of tensor calculus as the contraction between the gate $G_l^i$ and the environment tensor $\nabla^i_l f$ as described in Fig \ref{fig:tens_cost_func}.
To optimize $F_{\rm global}(\{G^l_i\})$ we go over all the gates in teh circuit, replacing $G^l_i$ with its optimal counterpart, the gate which maximizes the fidelity function. The process is repeated for many iterations until the cost function converges to a minimal value. Like before, here the optimal gate used at each step is the unitary closest to the matrix $\nabla^i_l f$ in trace distance, which can be found by the SVD algorithm in eq. \ref{eq:SVD}.

For the local cost function, the element by element optimization cannot be reduced into optimization of linear function of $G$, and so the optimization is done on terms quadratic in G, in the form of
\begin{equation}
\label{eq:local_quad_opt}
    F_{\rm local} = (G^\dagger)^{\Vec{o_2}}_{\Vec{i_2}}\; T^{\Vec{i_1},\Vec{i_2}}_{\Vec{o_1},\Vec{o_2}}\; G^{\Vec{o_1}}_{\Vec{i_1}},
\end{equation}
where $\Vec{i_1},\Vec{i_2},\Vec{o_1},\Vec{o_2}$ are the tensoric indices of $G$ and $G^\dagger$. 
In this case we can say that the term is a bi-linear form, and that $T$ is the second derivative of $F$ according to $G$ and $G^\dagger$.
Finding $G_{\rm optimal}$ in the bi-linear case is much harder, and an immediate analytical solution is hard to find.
Unlike optimization of bi-linear forms of normalized vectors, which can be calculated analytically for normalized vectors, here we require $G_{\rm optimal}$ to be unitary which results in a quadratic optimization problem under quadratic constraints.

Nevertheless, the optimal solution of eq. \ref{eq:local_quad_opt} can be calculated numerically by element-by-element sub-optimizations- replacing one copy of G with its optimal counterpart iteratively, which exhibits fast convergence rate to an optimized solution. Due to the symmetric structure of the local cost function, the optimal general solution for two arbitrary matrices is always symmetric as well, and so we can optimize $G$ and $G^\dagger$ independently and obtain a solution were the two matrices are the hermitian conjugate of one another.

\section{Gradient tomography for global cost functions}
\label{app:tomography}

In the main paper we discussed an optimization algorithm on a quantum hardware, where we calculate the fidelity and the cost function gradient of a quantum computer instead of classical simulations. To produce an efficient optimization algorithm, the evaluation of the gradient has to be translated into set of circuit measurement, in contrast to the tensor network approach in classical simulations. In particular, we would like to evaluate the gradient with respect to a full $n$-qubit unitary gate. Here we detail a process of gradient measurement which does not require angle parameterization, which we call gradient tomography. 

In gradient tomography we seek to reconstruct the gradient matrix of the cost function according to derivation along the single $n$-qubit gate $G^l_i$. The gradient can also be viewed as the "environment" tensor produced by the index contraction of all the circuit gates apart from $G^l_i$. Our approach is to perform a tomography of the environment tensor by choosing the unitary gate to be a tensor product of $n$ Pauli matrices.
Considering the case of global cost function when the overlap is  $f(G)=\mathrm{Tr}(T^\dagger G)$, the matrix $T$ can be projected onto different combinations of Pauli matrices 
\begin{align}
    t_{i_1,..i_n} = \frac{1}{2^n}\mathrm{Tr}( \sigma_{i_1}\sigma_{i_2}...\sigma_{i_n}T),
\end{align}

and combining the projections together using a sum of Pauli strings will reconstruct the tensor $T$,
\begin{equation}
    T = \sum_{i_k \in \{0,1,2,3\}}{t_{i_1,...,i_n} \sigma_{i_1}\sigma_{i_2}...\sigma_{i_n}}
\end{equation}
In practice, one can only measure $\abs{f(G)}^2$ using a quantum computer, and as a result the full reconstruction of $T$ is done in two parts. The magnitude of $t_{i_1,...,i_n}$ is obtain by choosing $G$ to be a tensor product of Pauli matrices and measuring $\abs{f(\sigma_{i_1}\sigma_{i_2}...\sigma_{i_n})}^2$, while the phases of $t_{i_1...i_n}$ can be extracted by measuring the relative phase between two Pauli terms
\begin{equation}
    \abs{f\left(\sigma_{i_1}\otimes...\frac{1}{\sqrt{2}}\left(\sigma_{i_j}+\zeta\sigma_{i'_j}\right)\otimes...\otimes\sigma_{i_n}\right)}^2 = \frac{1}{2}\abs{t_{i_1...i_n}}^2 + \frac{1}{2}\abs{t_{i_1...i'_j...i_n}}^2  + \re{t_{i_1...i_n}^*t_{i_1...i_j'...i_n}}
\end{equation}
from which the relative phase $\angle\left(t_{i_1...i_j+1...i_n}/t_{i_1...i_j...i_n}\right)$ can be isolated. The parameter $\zeta$ is either $1$ when both $\sigma_{i_j}$ and $\sigma_{i'_j}$ are proper pauli matrices with $i_j=1,2$ or $3$, and $1i$ when one of them is the identity matrix $\sigma_0$ , which keeps the unitarity of $G$ in either case.
In practice the tomography requires $2\times4^n - 1$ different substitution to recover $4^n$ amplitudes and $4^n - 1$ phases.
In comparison, using the parameter shift rule to calculate the full derivative according to every angle in the angle parameterization requires only $2\times4^n - 2$ measurements, when the extra measurement in the tomography process is related to a redundant additional degree of freedom in T which cancels out when projecting the environment onto the unitary manifold.
(Additionally one should note that a redundancy might appear in the phases reconstruction of $t_{i_1...i_n}$ due to the multiple solutions to the inverse of cosine, which may require additional measurements to resolve).
Despite the requirement of additional circuit measurements, our method presents a significant advantage over the phase shift rule method by requiring only application of 1-qubit gates on each qubit in order to characterize the full multi-qubit gradient. Extending this algorithm to local cost  functions presents more challenges, which  deserves further investigation.

\section{Additional simulation results}
\begin{figure}[h]
    \centering
    \includegraphics[width=300pt]{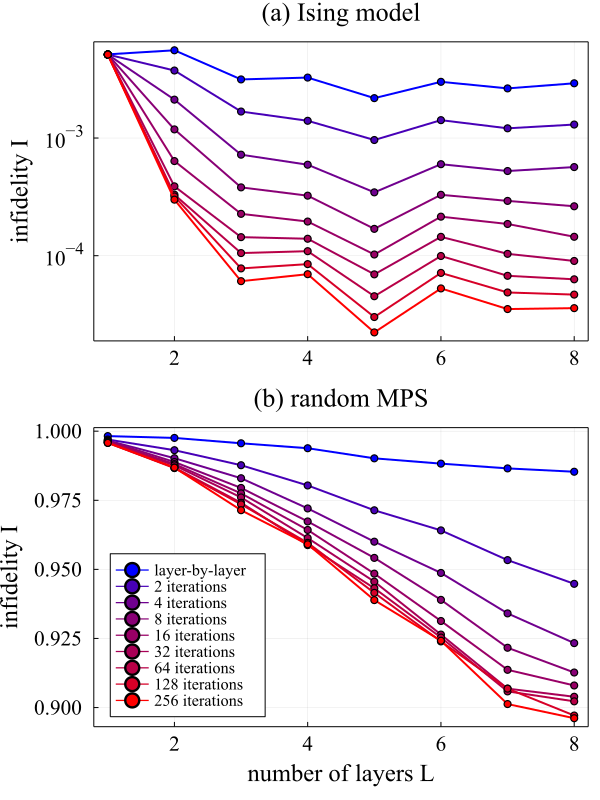}
    \caption{Same as Fig.~\ref{fig:infidelity} for $N=18$ qubits}
    \label{fig:18}
\end{figure}

\begin{figure}[h]
    \centering
    \includegraphics[width=300pt]{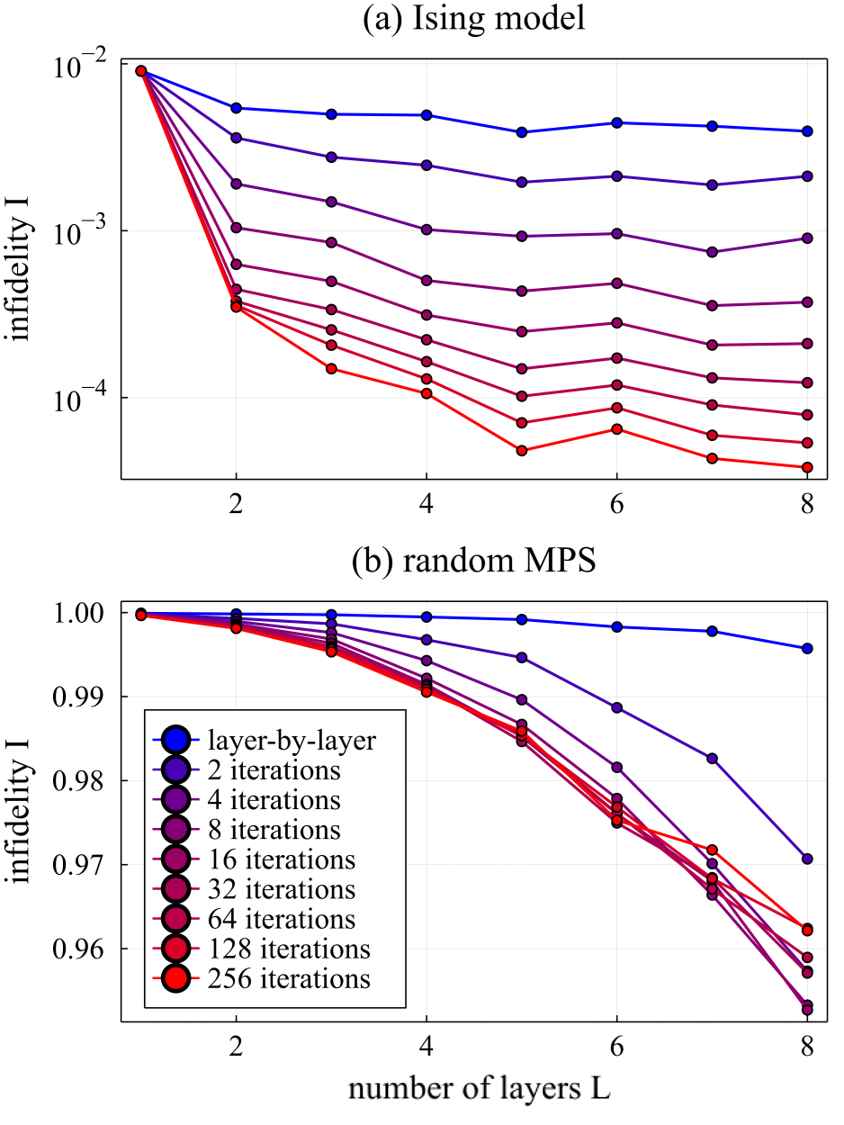}
        \caption{Same as Fig.~\ref{fig:infidelity} for $N=24$ qubits}
    \label{fig:24}
\end{figure}

Fig.~\ref{fig:18} and \ref{fig:24} show the infidelity of the optimization encoding obtained for random MPS target states with $N=18$ and $N=24$ qubits,  respectively. These results are consistent with Fig.~\ref{fig:infidelity} of the main text, with the only difference that in the case of a random MPS target state, the encoding delivers larger values of infidelity.

\label{sec:additional}
\newpage

\end{document}